\documentclass[11pt]{article}

\pdfoutput=1

\usepackage[makeroom]{cancel}
\usepackage{color}
\usepackage{graphicx}
\usepackage{amsmath}
\usepackage{amssymb}
\usepackage{xspace}
\usepackage[small]{subfigure}
\usepackage[numbers,compress]{natbib}
\usepackage[hyperfootnotes=false]{hyperref}
\usepackage{tcolorbox}

\newlength{\fighskip} \fighskip=2pt
\newlength{\figvskip} \figvskip=3pt

\newcommand*{\figbox}[2]{{
  \def\figscale{#1}
  \def\arraystretch{0.8}
  \arraycolsep=0pt
  \begin{array}{c}
    \vbox{\vskip\figscale\figvskip
      \hbox{\hskip\figscale\fighskip
        \includegraphics[scale=\figscale]{#2}}}
  \end{array}}}

\usepackage{mciteplus} 
\usepackage{dcolumn}
\usepackage{bm}
\usepackage{verbatim}
\usepackage{amscd}
\usepackage{amsfonts}
\usepackage{setspace}
\usepackage{amsthm}
\usepackage{enumerate}
\usepackage{mathtools}

\theoremstyle{plain}
\newtheorem{theorem}{Theorem}
\theoremstyle{plain}

\theoremstyle{plain}
 
\theoremstyle{plain}

\theoremstyle{remark}

\theoremstyle{conjecture}

\theoremstyle{observation}

\theoremstyle{definition}

\theoremstyle{corollary}

\theoremstyle{definition}

\theoremstyle{definition}

\theoremstyle{result}

\theoremstyle{assumption}

\theoremstyle{definition}

\theoremstyle{problem}

\theoremstyle{fact}

\DeclareMathOperator{\Tr}{Tr}

\begin{document}

\title{\bf 
Observer-Dependent Black Hole Interior \\
from Operator Collision
%Black Hole Interior from Operator Growth
}
\author{
Beni Yoshida\\ 
{\em \small Perimeter Institute for Theoretical Physics, Waterloo, Ontario N2L 2Y5, Canada} }
\date{}

\maketitle

\begin{abstract}

We present concrete construction of interior operators for a black hole which is perturbed by an infalling observer. The construction is independent from the initial states of the black hole while dependent only on the quantum state of the infalling observer. The construction has a natural interpretation from the perspective of the boundary operator's growth, resulting from the collision between operators accounting for the infalling and outgoing modes. The interior partner modes are created once the infalling observer measures the outgoing mode, suggesting that the black hole interior is observer-dependent. Implications of our results on various conceptual puzzles, including the firewall puzzle and the information problem, are also discussed. 
 
\end{abstract}

\tableofcontents

\newpage

\section{Introduction}

In the last decade our understanding of the black hole interior has significantly advanced by using ideas from quantum information theory. 
It has led to a conjecture that bulk operators within the entanglement wedge enclosed by the extremal surface of minimal area can be reconstructed on boundary degrees of freedom~\cite{Almheiri:2015aa, Pastawski15b, Dong2016, Jafferis:2016aa}. 
The attempt to find concrete expressions of bulk operators, however, has been only partially successful. 
While bulk operators in the smaller causal wedge were explicitly constructed~\cite{Hamilton06}, those outside the causal wedge remain elusive. 
Unfortunately black hole interior operators are the prime example of those behind the causal horizon. See a review~\cite{Harlow18} for a more complete list of references on this subject. 

Recently we pointed out that black hole interior operators can be constructed in boundary degrees of freedom by a procedure similar to the Hayden-Preskill thought experiment~\cite{Beni18, Beni19}. 
We envision that this observation is crucial in developing a generic prescription for reconstructing bulk operators behind the causal wedge and will provide important insights on the physics of the black hole interior. The goal of the present paper is to write down concrete expressions of interior operators by explicitly implementing the ideas outlined in these works. We also discuss their physical interpretations and implications to various conceptual puzzles concerning black hole interiors.

\begin{figure}
\centering
\includegraphics[width=0.45\textwidth]{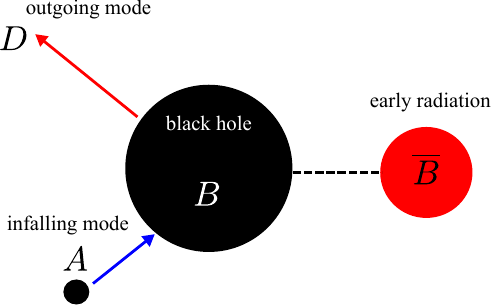}
\caption{A cartoon picture of the problem setting. A black hole is perturbed by an infalling observer. Our goal is to construct the interior partner operator of the outgoing mode. 
}
\label{fig-summary}
\end{figure}

The key challenge is to understand the experience of the infalling observer by examining the validity of semiclassical bulk descriptions properly. 
Useful hints can be obtained by considering the corresponding event in boundary degrees of freedom. What distinguishes our approach from many of previous ones is to include the infalling observer explicitly as a part of the quantum system (Fig.~\ref{fig-summary}). Adding an infalling observer has a rather dramatic effect which disentangles the outgoing mode from the early radiation completely. This conclusion follows from a certain quantum information theoretic relation, referred to as the decoupling theorem which asserts separability of two subsystems due to scrambling dynamics. Hence the preexisting entanglement plays no role in the infalling observer's experience. This is of course a counterargument against the ER=EPR approaches to the firewall puzzle~\cite{Maldacena13}.

The most salient feature of our construction is that it does not involve any degrees of freedom from the early radiation and hence is independent of initial states. This is a direct consequence of the aforementioned decoupling theorem. It is however important to recognize that the construction depends on how the infalling observer is introduced to the system and hence is observer-dependent. Observer-dependence of the black hole interior is reflected in the fact that expressions of interior modes are not unique, a manifestation of its quantum error-correction property~\footnote{It is perfectly fine to construct interior operators in a state-dependent manner by using the early radiation as in~\cite{Papadodimas:2013aa} for instance. However the decoupling theorem suggests that such construction is not relevant to the experience of the infalling observer when the backreaction is taken into account.}. 
We will see that observer-dependence provides resolutions and useful insights on various conceptual puzzles concerning black hole interior, including the firewall puzzle, the Page curve and the Marolf-Wall puzzle. 

Our construction bears some similarity with a pioneering work by Verlinde and Verlinde, written just 4 months after the AMPS paper, which utilized the idea of quantum error-correction to construct interior operators~\cite{Verlinde12}. The new ingredient in our work is to relate reconstructability to the concept of scrambling as diagnosed by out-of-time order correlators (OTOCs). Indeed, the construction is closely related to the boundary operator's growth in the AdS/CFT correspondence~\cite{Roberts:2015aa}. We will see that the interior partner of the outgoing operator $W$ can be constructed by time-evolving the infalling operator $V$ and looking at the overlap with $W$. This suggests that the interior modes are created once the outgoing modes are measured by the infalling observer, highlighting its observer-dependence.

Throughout the main text of the paper, we will use the infinite temperature approximation which treats a quantum black hole as an $n$-qubit quantum state where $S_{\text{BH}}=n$ is the Bekenstein-Hawking coarse-grained entropy. Generalization to finite temperature systems is discussed in appendix~\ref{sec:finite}. Unitarity and scrambling dynamics are assumed in this paper. We assume tensor factorization into subsystems for the Hilbert space, but we believe that this is not a crucial limitation of our argument.

This paper is organized as follows. In section~\ref{sec:decoupling}, we present the problem setting and the decoupling theorem. In section~\ref{sec:interior}, we obtain concrete expressions of interior partner operators. In section~\ref{sec:growth}, we relate the construction to the operator growth. In section~\ref{sec:observer-dependence}, we discuss observer-dependence. In section~\ref{sec:discussion}, we discuss implications of our construction to conceptual puzzles. In appendix~\ref{sec:review}, we present the definition of scrambling and summarize diagrammatic tensor notations. In appendix~\ref{sec:decoupling-proof}, we present a quantitative version of the decoupling theorem in terms of the $L^1$ distance. In appendix~\ref{sec:finite}, we obtain expressions of interior partner operators at finite temperature by imposing one technical assumption. In appendix~\ref{sec:calculation}, a certain calculation of OTOCs is presented. In appendix~\ref{sec:justification}, we justify the finite temperature assumption. 

\begin{figure}
\centering
\includegraphics[width=0.55\textwidth]{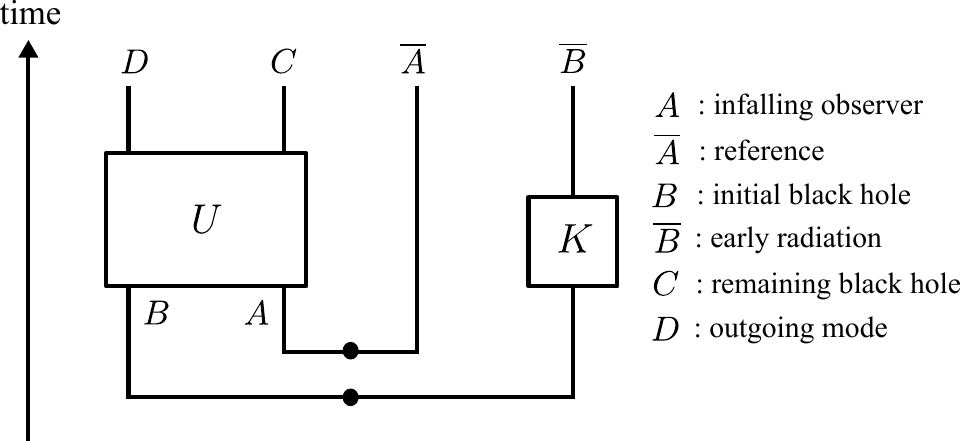}
\caption{A summary of the notation. 
}
\label{fig-notation}
\end{figure}

\section{Decoupling theorem}\label{sec:decoupling}

In this section we describe the problem setting which includes the infalling observer explicitly as a part of the quantum system. We then clarify the definition of interior partner operators. Finally we present the decoupling theorem which implies separability of the outgoing mode and the early radiation. A summary of our notation is given in Fig.~\ref{fig-notation} for convenience. 

\subsection{Black hole with infalling observer}

The initial state of the black hole is chosen to be a generic maximally entangled state between the initial black hole $B$ and the early radiation $\overline{B}$. Recall that a maximally entangled state can be represented as 
\begin{align}
(I \otimes K)|\text{EPR}\rangle_{B\overline{B}} \qquad \qquad |\text{EPR}\rangle_{B\overline{B}}\equiv\frac{1}{\sqrt{d_{B}}}\sum_{j}|j\rangle_{B}\otimes |j\rangle_{\overline{B}}
\end{align}
where $K$ is an arbitrary unitary operator acting on the early radiation $\overline{B}$. These qubits should be thought of as coarse-grained degrees of freedom of the black hole. 

Imagine that some measurement probe (or an infalling observer) $A$ is dropped into the black hole at time $t=0$. Depending on initial states of the probe, the black hole will respond in different manners due to the assumed unitarity of the dynamics. Rather than keeping track of the outcomes for all the possible input states on $A$, it is convenient to append a reference system $\overline{A}$ which is entangled with the probe $A$ and forming EPR pairs $|\text{EPR}\rangle_{A\overline{A}}=\frac{1}{\sqrt{d_{A}}}\sum_{j}|j\rangle_{A}\otimes |j\rangle_{\overline{A}}$. Being entangled with $A$, the reference $\overline{A}$ effectively keeps track of different input quantum states on $A$; if $\overline{A}$ is projected onto $|\psi^*\rangle$, $A$ is set to $|\psi\rangle$. The use of entangled reference system is a standard technique in black hole physics, dating back to the work by Hayden and Preskill~\cite{Hayden07}. 

After the time-evolution by scrambling dynamics $U$, the system evolves to
\begin{align}
|\Psi\rangle \equiv (U_{BA}\otimes I_{\overline{A}} \otimes K_{\overline{B}}) (|\text{EPR}\rangle_{B\overline{B}} |\text{EPR}\rangle_{A\overline{A}}) = \ \figbox{1.0}{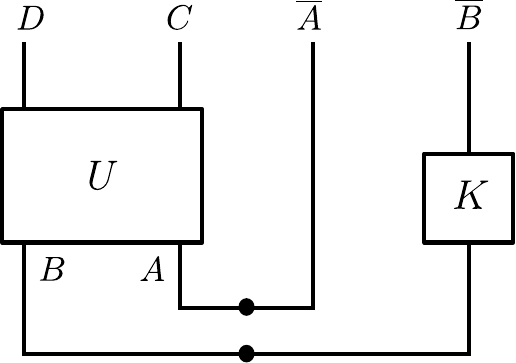} \label{eq:initial-state}
\end{align}
where $D$ is the outgoing mode and $C$ is the remaining black hole. Here black dots represent factor of $1/\sqrt{d_R}$ in a subsystem $R$ for proper normalization of $|\text{EPR}\rangle_{R\overline{R}}$. 

We are interested in constructing interior operators for this perturbed black hole state $|\Psi\rangle$. For an arbitrary operator $O_{D}$ on the outgoing mode, interior partner operators $\Omega(O_{D})$, supported on $C\overline{A}\overline{B}$, are defined to be those which satisfy 
\begin{align}
(O_{D}\otimes I_{C\overline{A}\overline{B}}) |\Psi\rangle \approx \big(I_{D}\otimes \Omega(O_{D})_{C\overline{A}\overline{B}} \big) |\Psi\rangle \label{eq:def-partner}
\end{align}
or graphically 
\begin{align}
\figbox{1.0}{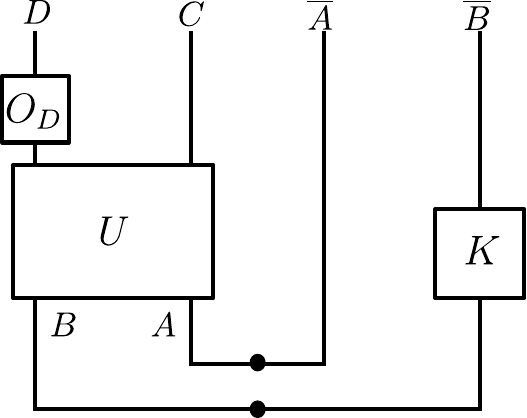} \ \approx \
\figbox{1.0}{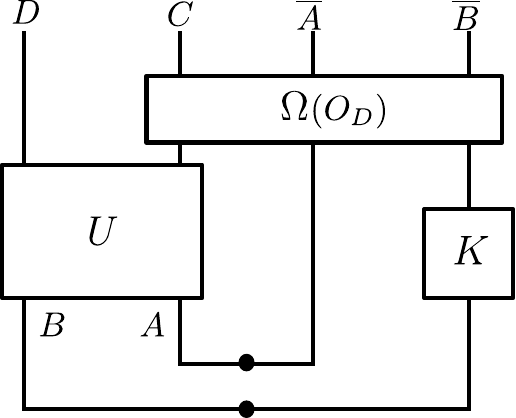}
\end{align}
Here ``$\approx$'' in Eq.~\eqref{eq:def-partner} is evaluated by the overlap between the LHS and RHS. 

The existence of partner operators is guaranteed since the mutual information $I(D,C\overline{A}\overline{B})$ is maximal, \emph{i.e.} $D$ is maximally entangled with $C\overline{A}\overline{B}$. 
One possible form of partner operators can be immediately constructed on $\overline{A}\overline{B}$ by evolving $O_{D}$ backwards in time and using the following relation
\begin{align}
(O\otimes I)|\text{EPR}\rangle = (I\otimes O^T)|\text{EPR}\rangle. \label{eq:EPR-relation}
\end{align}
This construction is state-dependent as it depends on the initial state $(I \otimes K)|\text{EPR}\rangle_{B\overline{B}}$ through its dependence on $K$.

\subsection{Decoupling theorem}

It cannot be emphasized enough, however, that construction of interior partners $\tilde{O}_{C\overline{A}\overline{B}}$ are not unique~\cite{Almheiri:2015aa}. The following (informally stated) decoupling theorem was proven in~\cite{Hosur:2015ylk, Yoshida:2017aa}. A more precise version is presented in appendix~\ref{sec:decoupling-proof}. 

\begin{theorem}\label{theorem:decoupling}
(Informal) Suppose that the system is scrambling in a sense of Eq.~\eqref{eq:scrambling-definition} and $d_A\gg d_D$. Then, the subsystems $D$ and $\overline{B}$ are decoupled (not entangled):
\begin{align}
\rho_{D\overline{B}} \approx \rho_{D} \otimes \rho_{\overline{B}}  \label{eq:decoupling}.
\end{align}
where the error is $O\big(\frac{d_{D}^2}{d_{A}^2}\big)$.
Furthermore, for an arbitrary operator $O_{D}$ on the outgoing mode $D$, a partner operator $\Omega(O_{D})_{C\overline{A}}$ can be constructed on $C\overline{A}$ without using any degrees of freedom from the early radiation $\overline{B}$. 
\end{theorem}

We present a brief review of the definition of scrambling in appendix~\ref{sec:review}. Loosely speaking it says that OTOCs decay to small values. This is a universal property of quantum black holes whose background geometries have a well-defined horizon. 

Note that in the absence of $A$, the outgoing mode $D$ would be entangled with the early radiation. 
The theorem states that the outgoing mode $D$ is no longer entangled with the early radiation $\overline{B}$ due to the perturbation. Equivalently, the outgoing mode $D$ is (almost) maximally entangled with the remaining black hole $C$ and the reference $\overline{A}$. 

It is useful to represent the partner operator from theorem~\ref{theorem:decoupling} graphically as follows:
\begin{align}
\figbox{1.0}{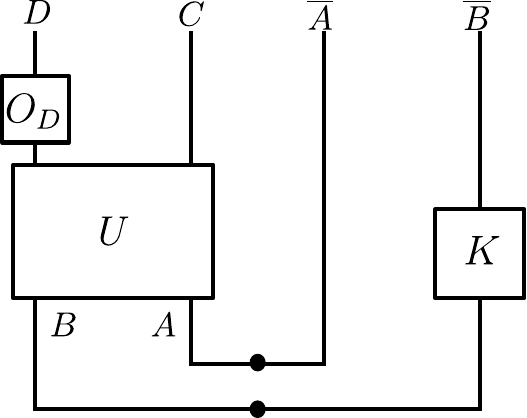} \ \approx \
\figbox{1.0}{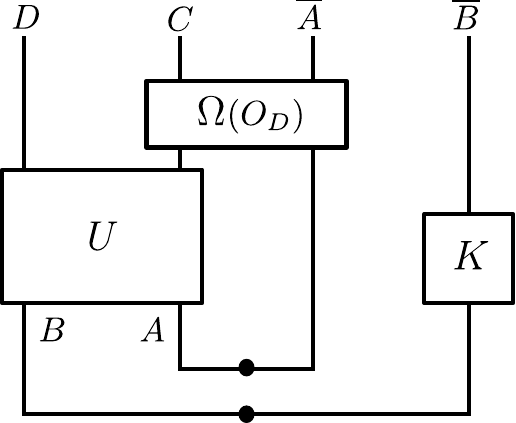} \ .
\end{align}
Such construction $\Omega(O_{D})$ must be state-independent as it does not depend on the unitary $K$. In the next section we construct the interior operator $\Omega(O_{D})$ explicitly.

\section{Interior operator}\label{sec:interior}

Our construction of interior operators utilizes a certain quantum operation which distills EPR pairs between $D$ and $C\overline{A}$. The procedure is inspired by a probabilistic recovery protocol for the Hayden-Preskill thought experiment~\cite{Yoshida:2017aa}. A deterministic version of the recovery protocol from~\cite{Yoshida:2017aa} works equally well for this purpose, but we will focus on the probabilistic one for simplicity. 

\subsection{Entanglement distillation}

Given the time-evolved state $|\Psi\rangle$ in Eq.~\eqref{eq:initial-state}, let us construct the following state:
\begin{align}
|\Phi_{\text{back}}\rangle \equiv \ \figbox{1.0}{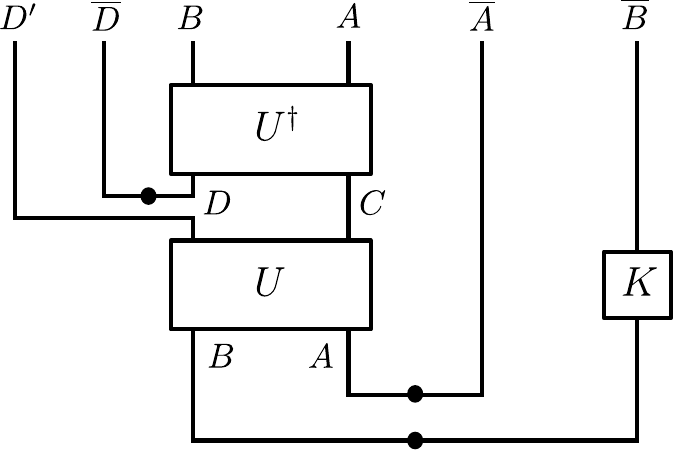}\ .
\end{align}
Here ancillary subsystems $D'\overline{D}$ are introduced, qubits on $D$ are moved to $D'$, $|\text{EPR}\rangle_{\overline{D}D}$ is prepared on $\overline{D}D$, and the time-reversal $U^{\dagger}$ is applied on $DC$.  Time-reversal $U^{\dagger}$ is unphysical in a real world, but this is just a mathematical trick to obtain the interior construction.

So far all the operations are deterministic. We now apply a projection onto EPR pairs on $A\overline{A}$. The unnormalized outcome is 
\begin{align}
\Pi^{(\text{EPR})}_{A\overline{A}} |\Phi_{\text{back}}\rangle = \ \figbox{1.0}{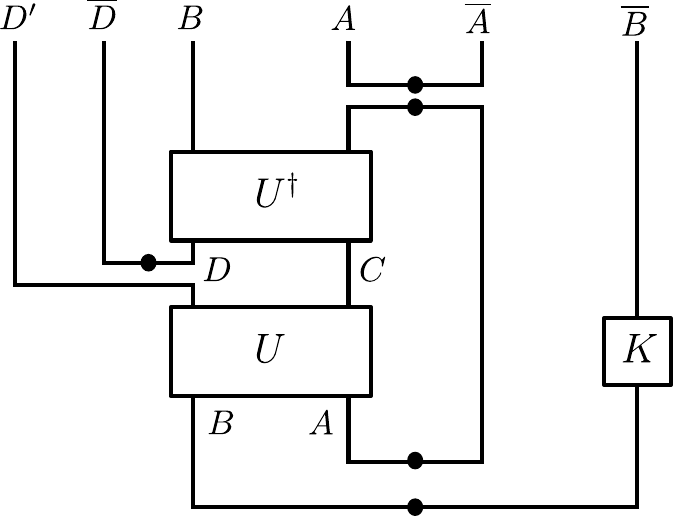}
\end{align}
where the projector is denoted by $\Pi^{(\text{EPR})}_{A\overline{A}}$.
The probability amplitude for this event is
\begin{align}
\Delta \equiv \langle \Phi_{\text{back}}|\Pi^{(\text{EPR})}_{A\overline{A}} |\Phi_{\text{back}}\rangle = \ \figbox{1.0}{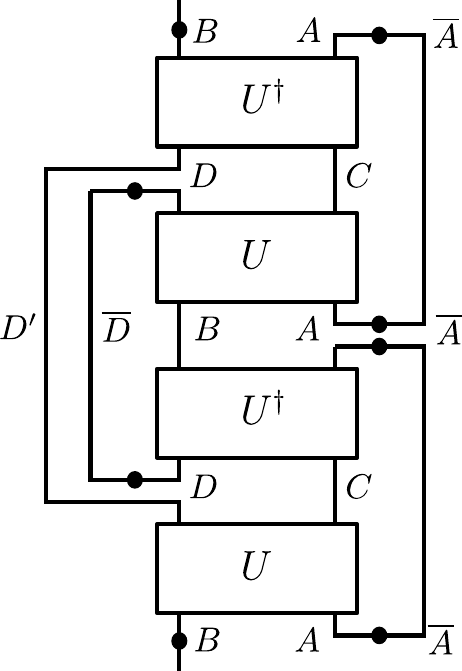}\ .
\end{align}
A line connecting two $B$'s on the top and bottom of the diagram is omitted for ease of drawing.
Finally the normalized outcome after the postselection is given by 
\begin{align}
|\widetilde{\Phi}_{\text{back}} \rangle  \equiv \frac{1}{\sqrt{\Delta}} \Pi^{(\text{EPR})}_{A\overline{A}}|\Phi_{\text{back}}\rangle. 
\end{align}

Some readers may have already noticed a relation between $\Delta$ and OTOCs by staring at the diagrammatic representation of $\Delta$ where $U$'s and $U^{\dagger}$'s appear twice. The relation between OTOCs and $\Delta$ can be explicitly seen by the following identity~\cite{Hosur:2015ylk}
\begin{align}
\Delta = \iint \langle O_{A}(0) O_{D}(t) O_{A}^{\dagger}(0) O_{D}^{\dagger}(t) \rangle d O_{A}dO_{D}.
\end{align}
where the integrals over $O_{A}$ and $O_{D}$ represents uniform averaging over all the basis operators on $A$ and $D$ respectively. Hence, the amplitude $\Delta$ can be evaluated by using the asymptotic form of OTOCs for scrambling systems in Eq.~\eqref{eq:scrambling-definition} in appendix~\ref{sec:review}. After some calculation, we obtain
\begin{align}
\Delta = \frac{1}{d_{D}^2} \left[ 1 + O\left( \frac{d_{D}^2}{d_{A}^2}\right) \right] \qquad (d_{D}\ll d_{A}).
\end{align}

We now show that EPR pairs have been already distilled on $D'\overline{D}$ after the postselection by $\Pi^{(\text{EPR})}_{A\overline{A}}$. The probability of observing EPR pairs on $D'\overline{D}$ is given by 
\begin{equation}
\begin{split}
 \langle \widetilde{\Phi}_{\text{back}} | \Pi^{(\text{EPR})}_{D'\overline{D}} | \widetilde{\Phi}_{\text{back}} \rangle 
 &= \frac{1}{\Delta}\langle \Phi_{\text{back}} | \Pi^{(\text{EPR})}_{A\overline{A}}\Pi^{(\text{EPR})}_{D'\overline{D}} | \Phi_{\text{back}} \rangle \\
 &= \frac{1}{\Delta}\frac{1}{d_{D}^2} \\
 &\approx 1. \label{eq:distillation}
\end{split}
\end{equation}
In the second step, we have used the following identity 
\begin{align}
\langle \Phi_{\text{back}} | \Pi^{(\text{EPR})}_{A\overline{A}}\Pi^{(\text{EPR})}_{D'\overline{D}} | \Phi_{\text{back}} \rangle= \ \figbox{1.0}{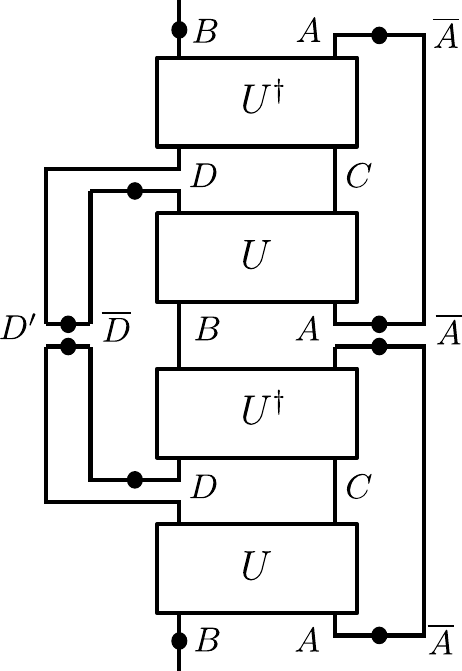} \  = \frac{1}{d_{D}^2}.
\end{align}

\subsection{Construction of interior operator}

To construct interior partner operators $\Omega(O_{D})$ for the perturbed black hole state $|\Psi\rangle$, we need to undo the projection $\Pi_{A\overline{A}}$ and the time-reversal $U^{\dagger}$. Partner operators are explicitly given by
\begin{align}
\Omega(O_{D}) \equiv  \frac{1}{\Delta}\frac{1}{d_{D} }\Tr\Big[ \big(O_{D}\otimes I_{C\overline{A}}\big) \big(U_{BA}\otimes I_{\overline{A}}\big) \big( I_{B}\otimes \Pi^{\text{(EPR)}}_{A\overline{A}}\big) \big(U^{\dagger}_{BA} \otimes I_{\overline{A}}\big) \Big]
\end{align}
or graphically 
\begin{align}
\Omega(O_{D}) \equiv  \frac{1}{\Delta} \  \figbox{1.0}{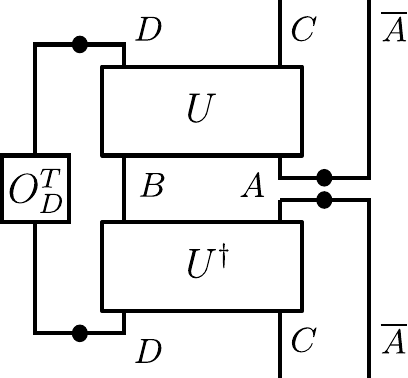} \ . \label{eq:construction}
\end{align}

Most readers will not find this expression particularly illuminating. In the next section we will further simplify the expression and discuss its relation to the boundary operator's growth. 

It can be verified that these are appropriate partner operators by evaluating the overlap between $(O_{D}\otimes I)|\Psi\rangle$ and $(I\otimes \Omega(O_{D}))|\Psi\rangle$:
\begin{align}
\langle \Psi | (O_{D}^{\dagger}\otimes \Omega(O_{D}))|\Psi \rangle = \   \figbox{1.0}{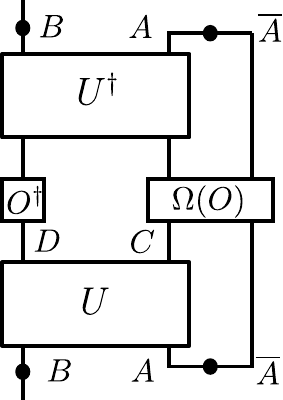} = \frac{1}{\Delta} \   \figbox{1.0}{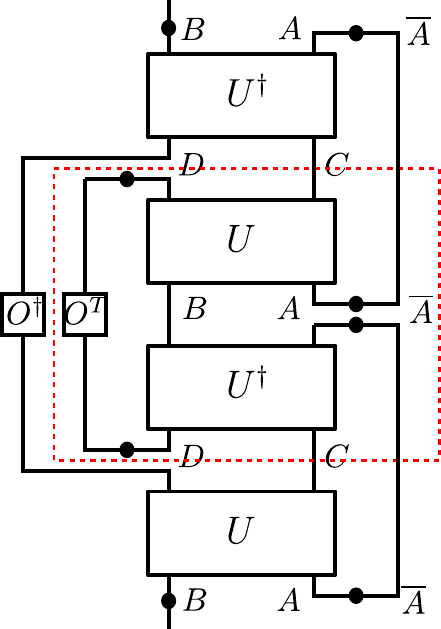}
\end{align}
where a part enclosed by a dotted square corresponds to $\Omega(O_{D})$. Taking average over basis operators on $D$, we have 
\begin{align}
\int d O_{D} \langle \Psi | (O_{D}^{\dagger}\otimes \Omega(O_{D}))|\Psi \rangle = \frac{1}{\Delta}\langle \Phi_{\text{back}} | \Pi^{(\text{EPR})}_{A\overline{A}}\Pi^{(\text{EPR})}_{D'\overline{D}} | \Phi_{\text{back}} \rangle = \frac{1}{\Delta}\frac{1}{d_{D}^2} \approx 1 \label{eq:overlap}
\end{align}
where we used $\int d O\ O \otimes O^{*}  = |\text{EPR}\rangle\langle \text{EPR}|$. Hence $\langle \Psi | (O_{D}^{\dagger}\otimes \Omega(O_{D}))|\Psi \rangle\approx 1$. 

It is worth emphasizing that our construction has a concrete operational meaning of distilling EPR pairs between the outgoing mode $D$ and the remaining black hole $C$ plus the reference $\overline{A}$. Indeed the average overlap in Eq.~\eqref{eq:overlap} is equal to the EPR distillation probability in Eq.~\eqref{eq:distillation}. 

\subsection{Pure state input}

Instead of introducing a reference system $\overline{A}$ entangled with $A$, let us consider a scenario where a measurement probe $A$, prepared in a particular pure state (e.g. $|0\rangle$), is dropped toward the black hole. 

The procedure to distill EPR pairs and construct interior operators proceeds in an analogous manner except that this time we need to project $A$ onto $|0\rangle$:
\begin{align}
\Pi_{A}^{|0\rangle\langle 0|} |\Phi_{\text{back}}\rangle = \  \figbox{1.0}{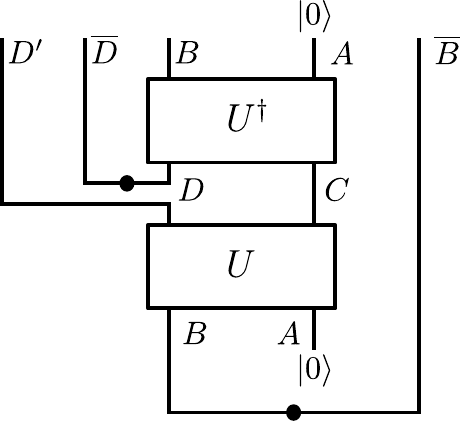} \ . \label{eq:construction-infalling}
\end{align}
The probability amplitude for the projection onto $|0\rangle$ can be computed from the OTOC asymptotics: 
\begin{align}
\Delta = \ \figbox{1.0}{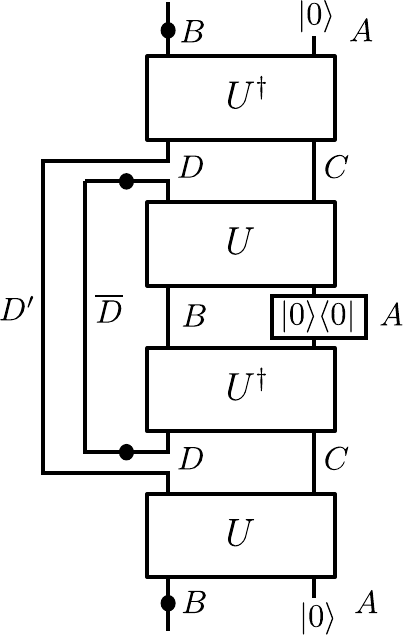} \ \approx \frac{1}{d_{D}^2} \qquad (d_{A}\gg d_{D}^2).
\end{align}
Note that the reconstruction criteria becomes $d_{A}\gg d_{D}^2$ instead of $d_{A}\gg d_{D}$. 

The probability of observing EPR pairs on $D'\overline{D}$ is $\frac{1}{\Delta} \langle \Phi_{\text{back}}| \Pi_{A}^{|0\rangle \langle 0|} \Pi_{D'\overline{D}} |\Phi_{\text{back}}\rangle = \frac{1}{\Delta}\frac{1}{d_{D}^2} \approx 1$. Interior partner operators are given by
\begin{align}
\Omega(O_{D}) \equiv  \frac{1}{\Delta} \  \figbox{1.0}{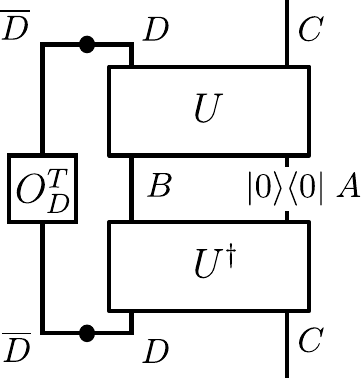} \ . \label{eq:construction-state}
\end{align}

\section{Operator growth interpretation}\label{sec:growth}

In this section we will relate the aforementioned construction to the operator growth interpretation of scrambling~\cite{Roberts:2015aa}. 

The operator growth picture relies crucially on the finiteness of the Hilbert space. It is an interesting future problem to generalize our construction to quantum field theories. See~\cite{Zhuang:2019aa} for discussions on oscillator systems. 

\subsection{Truncation by local operator}

Scrambling phenomena can be associated with the boundary operator's growth. Given a single qubit operator $V$ in a system of $n$ qubits, consider its time evolution by the Hamiltonian $V(t)=e^{-iHt}V e^{iHt}$:
\begin{align}
V(t) = V - iHt [H,V] - \frac{t^2}{2!} [H,[H,V]] + \frac{i t^3}{3!}[H, [H,[H,V]]] +\cdots.
\end{align}
Commutators $[H,\cdots [H,[H,V]]] $ generically increases the number of supports and $V(t)$ becomes a high weight operator for large $t$. The growth of $V(t)$ can be probed by OTOCs. When $W$ is some other single qubit operator on a different qubit and $V(t)$ has nontrivial support on that qubit, $\langle V(t) W(0) V(t) W(0)  \rangle$ decays from its initial value. 

The operator growth can be related to interior reconstruction by considering \emph{truncation} of operators. For an arbitrary operator $V$ on a Hilbert space $\mathcal{H} = \mathcal{H}_{C}\otimes \mathcal{H}_{D}$, its truncation by $W_{D}$ is defined by
\begin{align}
V_{/ W_{D}} \equiv \frac{1}{d_{D}} I_{D}\otimes  \Tr[ W_{D}^{\dagger} V ] = \ \figbox{1.0}{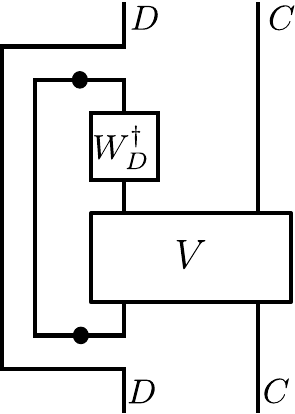}\ .
\end{align}
Note that truncated operator $V_{/ W_{D}} $ acts non-trivially only on $C$. In the Choi representation $| O \rangle \equiv (O\otimes I)|\text{EPR}\rangle$, the truncation can be defined by $| V_{/W_{D}} \rangle \equiv | I_{D}\rangle \langle W_{D}| V \rangle $. 

To gain some intuition, let us expand $V$ in terms of Pauli operators: $V = \sum_{P \in \text{Pauli}} \alpha_{P} P$. 
Assume that $D$ consists of just a single qubit and choose $W_{D}$ to be a Pauli operator $\{I_{D},X_{D},Y_{D},Z_{D} \}$. Define four sets of Pauli operators on $\mathcal{H}$, depending on the Pauli operator supported on $D$:
\begin{align}
\text{Pauli}|_{P_{D}} &= \{P_{D} \otimes \text{Pauli Operators on $C$}\} \qquad P_{D}=I_{D},X_{D},Y_{D},Z_{D}.
\end{align} 
The truncation $V_{/ W_{D}}$ is constructed by picking up terms with Pauli operators containing $W_{D}$ on $D$:
\begin{align}
V_{/ W_{D}} = \sum_{P \in \text{Pauli}|_{W_{D}}} \alpha_{P}P.
\end{align}
In other words, the truncation give rise to non-trivial operators when the original operator $V$ and the target operator $W_{D}$ collide with each other.

\subsection{Interior operator from collision}

The notion of truncation simplifies the expression of interior operators. Using 
\begin{align}
\int d O_{A} \ O_{A} \otimes O_{\overline{A}}^* = |\text{EPR}\rangle \langle \text{EPR}|_{A\overline{A}}, 
\end{align}
the interior partner operator in Eq.~\eqref{eq:construction} can be expressed concisely as
\begin{align}
\Omega(O_{D}) = \frac{1}{\Delta} \int d O_{A} \ \figbox{1.0}{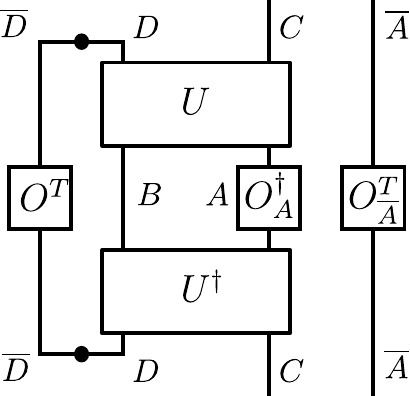}\ = \ \frac{1}{\Delta} \int dO_{A} \ O_{A}^{\dagger}(-t)_{/ O_{D}^{\dagger}} \otimes O_{\overline{A}}^T
\end{align}
with the integral over basis operators on $A$. The reason why the backward-evolution $O_{A}(-t)$ appears can be understood by rewriting OTOCs as $\langle O_{D}^{\dagger}(t) O_{A}^{\dagger}(0)O_{D}(t)O_{A}(0) \rangle=\langle  O_{D}^{\dagger}(0)O_{A}^{\dagger}(-t) O_{D}(0)O_{A}(-t) \rangle$ which evaluates the overlap between $O_{A}(-t)$ and $O_{D}(0)$.

Some readers may feel uncomfortable about taking an average over $O_{A}$. When $A$ is sufficiently large, we may pick one Pauli operator $O_{A}$ at random to construct a partner operator $\frac{1}{\Delta} O_{A}^{\dagger}(-t)_{/O_{D}^{\dagger} } \otimes O_{\overline{A}}^{T} $ without taking averages since statistical variances are suppressed in a scrambling system. The relation between the operator growth and interior operator construction becomes transparent; when $O_{A}(-t)$ collides with the target outgoing operator $O_{D}$, interior partner operators can be constructed. 

For the cases with a pure state input $|0\rangle$ on $A$, we obtain a simpler expression:
\begin{align}
\Omega(O_{D}) = \frac{1}{\Delta} O_{A}(-t)_{/ O_{D}^{\dagger}} \qquad O_{A} = I_{B}\otimes |0\rangle\langle 0 |_{A}.
\end{align}
The physical interpretation is clearer in this case. Recall that $|0\rangle$ was the initial state of the infalling observer and $O_{A}$ is its projection. Hence the interior partner operators are constructed due to the collision between the infalling observer and the target outgoing $O_{D}$. This suggests that the interior modes are created once the outgoing modes are measured by the infalling observer.

\subsection{On SYK model}

Finally we make a brief comment on interior operators for the SYK model. Since the SYK model exhibits scrambling dynamics, our prescription can be applied to obtain the entangled partner. It has a simple Feynman diagrammatic interpretation modulo one technical subtlety. 

\begin{figure}
\centering
\includegraphics[width=0.35\textwidth]{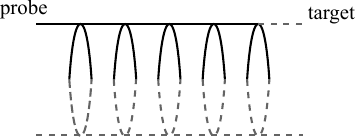}
\caption{The partner operator as the upper half of the ladder diagram ($q=4$ SYK). 
}
\label{fig-cartoon2}
\end{figure}

Consider the OTOC $\langle \psi_1(0) \psi_N (t) \psi_{1}(0) \psi_N(t) \rangle$ where $\psi_1,\psi_N$ are interpreted as infalling and outgoing modes respectively. Here we would like to construct a partner of $\psi_N$. The dominant contribution to OTOCs comes from the ladder diagram, see Fig.~\ref{fig-cartoon2}. In~\cite{Roberts:2018aa}, it has been pointed out that the ladder diagram characterizes the collision between $\psi_1(-t)$ and $\psi_{N}(0)$. In this interpretation, the upper half of the ladder diagram can be viewed as the dominant contribution in the growth of $\psi_1(-t)$. Hence, by truncating $\psi_N$ from the upper half of the ladder diagram, we obtain the interior operator (Fig.~\ref{fig-cartoon2}).

It should be however noted that the above diagrammatic interpretation has a problem. The ladder diagram account for the dominant piece in the OTOC decay until the scrambling time. Around the scrambling time, the sub-leading terms become important. Since the fidelity of our reconstruction becomes better when OTOCs have decayed to small values, the diagrammatic expression in Fig.~\ref{fig-cartoon2} is a good description of interior partner operator only for some time window. 

\section{Observer-dependent interior}\label{sec:observer-dependence}

In this section we discuss state-independence and observer-dependence of interior operators. See~\cite{Harlow:2014aa} for a list of references and possible problems associated with state-dependence.

\subsection{State-independence}

So far we have considered maximally entangled black holes whose initial states are given by $(I\otimes K)|\text{EPR}\rangle_{B\overline{B}}$ with an arbitrary unitary $K$. The construction of interior operators $\Omega(O_{D})_{C\overline{A}}$ is state-independent in a sense that it does not depend on the initial state of the black hole. 

The same construction works for pure state black holes too since its expression does not have any non-trivial support on the early radiation $\overline{B}$. To see this explicitly, let us project $\overline{B}$ onto an arbitrary pure state $|\psi_{\text{BH}}^*\rangle$. This amounts to considering the initial state of the form
\begin{align}
\figbox{1.0}{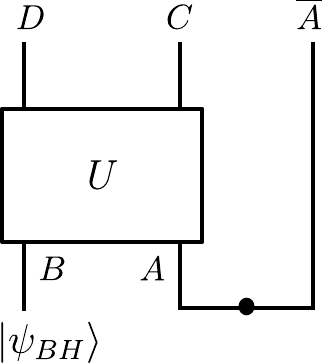}\ . 
\end{align}
Repeating the same calculation by replacing $K$ with $|\psi_{\text{BH}}^*\rangle\langle \psi_{\text{BH}}^*|$, we can easily verify that the same expression $\Omega(O_{D})_{C\overline{A}}$ still satisfies the criteria of interior partner operators in Eq.~\eqref{eq:def-partner}. 

The fact that the same construction $\Omega(O_{D})_{C\overline{A}}$ works universally for all the black hole initial states has somewhat counterintuitive implication. This suggests that how the outgoing mode $D$ is entangled with $C\overline{A}$ is a universal property of the scrambling dynamics $U$ and is independent of the initial state in $B$, be it pure or mixed. 
It is interesting that strongly chaotic dynamics generates robust quantum correlations which are insensitive to initial conditions !

The state-independence is a direct implication of the decoupling theorem~\ref{theorem:decoupling} which says the outgoing mode $D$ is disentangled from the early radiation $\overline{B}$. Due to this decoupling phenomena, preexisting entanglement between the black hole and the early radiation plays no important role for the experience of the infalling observer. The bulk explanation of the decoupling phenomena and its implication on the firewall puzzle are discussed in section~\ref{sec:discussion}. 

\subsection{Observer-dependence}

We have seen that the black hole interior operator can be constructed in a way independent of the initial state due to the infalling observer's backreaction. It is however important to recognize that the construction is dependent on how the infalling observer $A$ is introduced to the system. Hence the construction is state-independent but is observer-dependent. This suggests that the black hole interior should be considered as observer-dependent. 

In fact we have already seen observer-dependent nature of the black hole interior rather vividly in the operator growth interpretation. We found that the interior operator can be constructed by examining the collision between the outgoing mode $W$ and the infalling mode $V(-t)$. The implication is that the interior mode was created dynamically once the infalling observer measures the outgoing mode. This interpretation further hints observer-dependence of the black hole interior.

We may formalize the notion of observer-dependence by borrowing a quantum information theoretic language. Let us consider a maximally entangled black hole and recall how the interior modes are realized for the outside and infalling observers respectively. In the absence of the infalling observer, the entangled partner of the outgoing mode can be found as some degrees of freedom in the early radiation. This is what the outside observer perceives as the interior mode. Yet, from the perspective of the infalling observer, this thing in the early radiation is not the interior mode anymore since it has been already decoupled from the outgoing mode. The infalling observer will perceive the mode which we have identified in previous sections as the true interior partner mode.  

As is evident from this argument, observer-dependence of the black hole interior is closely related to the fact that expressions of interior partner operators are not unique. This property can be better understood by using an analogy with quantum error-correcting codes. In the language of quantum error-correcting code, partner operators can be interpreted as logical operators where quantum information on the outgoing mode $D$ is encoded into $C\overline{AB}$. Consider the following embedding map $\Gamma : \mathcal{H}_{D} \rightarrow \mathcal{H}_{\text{code}}$ defined by
\begin{align}
|\psi\rangle \in \mathcal{H}_{D} \ \overset{\Gamma} {\longrightarrow} \ \Gamma(|\psi\rangle) = \langle \psi |_{D} |\Psi\rangle_{DC\overline{AB}} \in \mathcal{H}_{\text{code}} \subseteq \mathcal{H}_{C\overline{AB}}
\end{align}
where we omitted the normalization factor for simplicity. For readers familiar with tensor diagrammatic notations, it is convenient to represent the embedding map graphically as follows by rotating the original diagram for $|\Psi\rangle$ 180 degree:
\begin{align}
\figbox{1.0}{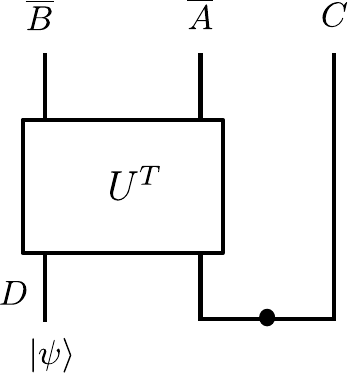}
\end{align}
Here partner logical operators are those which may transform wavefunctions in the code subspace $\mathcal{H}_{\text{code}}$ in a non-trivially manner. In the quantum error-correction interpretation, logical operators on $\overline{A}\overline{B}$ is what the outside observer perceives as the entangled partner whereas those on $C\overline{A}$ is what the infalling observer will measure when approaching the horizon. 

We have argued that quantum error-correcting nature of interior reconstruction gives rise to observer-dependence. At this point, it is interesting to recall that the holographic mapping from bulk operators to boundary operators in the AdS/CFT correspondence can be understood as a quantum error-correcting code. The canonical example of holographic quantum error-correction is the empty AdS space where the bulk operator at the center can be reconstructed on any interval which is longer than the half of the system via the AdS/Rindler reconstruction. Our interpretation is that each boundary expression corresponds to what the boundary observer, who has an access to the given boundary interval, perceives as the bulk operator. 

Observer-dependence of a black hole itself is a natural idea from the perspective of its thermodynamic property. For instance the no hair theorem tells that black holes in thermal equilibrium are characterized by a few macroscopic observables. This suggests that whatever happened in the past will not affect the present observer as far as outgoing modes are concerned. A typical time scale for this is order of the thermal time. 

What we advocate here is distinct from this conventional wisdom based on the thermodynamics of black holes. We argue that the black hole interior, which is understood as entangled partners of outgoing modes, must be observer-dependent too. A typical time scale for this is order of the scrambling time. 

We conclude this section by recalling a certain old argument by Hayden and Preskill~\cite{Hayden07}. Let us consider two observers who jump into the one-sided black hole and try to meet inside the black hole. If the time separation $\Delta t$ between them is longer than the scrambling time, an observer needs large amount of energy, larger than the Planck energy, to send a signal to the other observer. Hence two observers will not be able to meet. This was the crucial observation by Hayden and Preskill which led to the lower bound on the scrambling time. 

Our interpretation of their argument is as follows. When two observers are separated by more than the scrambling time, the black hole interiors they observe are unrelated. They can enjoy their own interiors without being perturbed by others. 

\section{Discussions}\label{sec:discussion}

We have presented concrete expressions of the entangled partner operators in boundary degrees of freedom for a quantum black hole perturbed by an infalling mode. The primary assumption in our derivation was that the black hole scrambles quantum information in a sense of OTOC decay. 
Construction becomes possible due to the overlap between operators accounting for the infalling and outgoing modes. 
This collision disentangles the outgoing mode from the early radiation and creates a new entangled partner mode as stated in the decoupling theorem. 
The important implication is that interior operators, perceived by the infalling observer are state-independent, but are observer-dependent. 

We have not discussed why we can take $d_{A}\gtrapprox d_{D}$. One possible explanation is that in order to experience some physics, the observer herself should carry some amount of entropies, at least as much as the objects she is going to measure. In fact, this naive argument is enough to resolve the firewall puzzle, namely the apparent contradiction demonstrated in the AMPS thought experiment as we shall see below. We are however not very satisfied with this metaphysical explanation. 

Perhaps a better question to ask is whether we actually need to include the infalling observer as additional degrees of freedom or not in deriving the bulk description from the boundary description. 
From the outside viewpoint, the infalling and outgoing modes influence each other in a dynamical manner. 
When switching to the infalling observer's viewpoint, infalling modes somehow decohere, perhaps due to the choice of the infalling trajectory.
Ultimately we hope to understand why the interior operator is automatically decoded simply by switching to the infalling observer's frame.

In the reminder of this section, we present discussions on several puzzles concerning the black hole interior. 

\subsection{Firewall puzzle}\label{sec:firewall}

The aforementioned reconstruction of interior partner operators has prompted us to propose a possible resolution of the firewall problem~\cite{Beni19} (in a sense of the monogamy puzzle~\cite{Mathur:2009aa, Braunstein:2013aa, Almheiri13}). 

\begin{figure}
\centering
\includegraphics[width=0.45\textwidth]{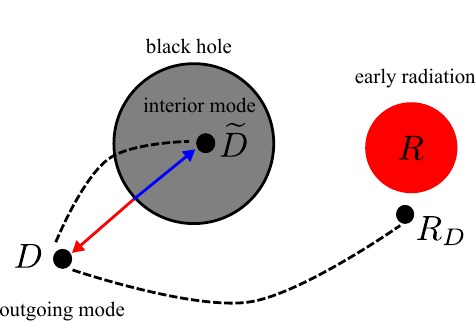} 
\caption{Summary of the firewall argument. The outgoing mode $D$ cannot be entangled with the early radiation $R$ and the interior mode $\widetilde{D}$ simultaneously. 
}
\label{fig-firewall}
\end{figure}

Let us briefly recall the firewall argument (Fig.~\ref{fig-firewall}). Given an old black hole $B$ which is maximally entangled with the early radiation $R$, let us consider an outgoing mode $D$ which was just emitted from the black hole. Since the black hole is maximally entangled, the outgoing mode $D$ must be entangled with some degrees of freedom in $R$. Then the outside observer, Bob, may distill a qubit $R_{D}$ from $R$ that is entangled with the outgoing mode $D$ and hand it to the infalling observer Alice who is going to jump into a black hole. After crossing the smooth horizon, Alice will see an interior mode $\widetilde{D}$ which is entangled with the outgoing mode $D$. This, however, leads to a contradiction because $D$ is also entangled with $R_{D}$. If we believe what Bob says, $D$ has to be entangled with $R_{D}$, so $D$ and $\widetilde{D}$ are not entangled, leading to possible high energy density at the horizon.

The resolution of the firewall puzzle is immediate from a physical interpretation of the aforementioned reconstruction of the interior partner operators. When Alice falls into a black hole, the outgoing mode $D$ is disentangled from the distilled qubit $R_{D}$ due to the backreaction which makes OTOCs decay. In fact, the decoupling theorem~\ref{theorem:decoupling} tells that $D$ is decoupled from any degrees of freedom in the early radiation $R$. Hence, the monogamy of quantum entanglement is not violated. Alice will be able to observe the interior mode $\widetilde{D}$ which is distinct from the original partner mode $R_{D}$.

\begin{figure}
\centering
(a)\includegraphics[width=0.35\textwidth]{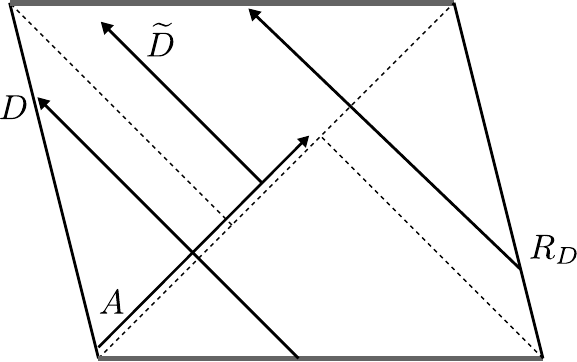} 
(b)\includegraphics[width=0.265\textwidth]{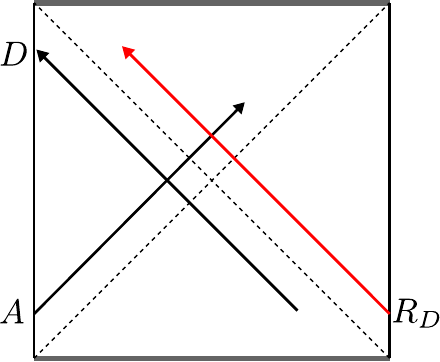} 
\caption{(a) The bulk interpretation of the proposed resolution of the firewall puzzle. The infalling observer $A$ is treated as a gravitational shockwave. (b) A schematic picture of previous proposals. This is not consistent with unitarity or scrambling of the black hole dynamics.
}
\label{fig-Penrose}
\end{figure}

The bulk interpretation of the aforementioned resolution can be obtained by treating the backreaction from Alice as a gravitational shockwave (Fig.~\ref{fig-Penrose}(a)). For simplicity of drawing, let us consider the AdS eternal black hole. Given the outgoing mode $D$, one possible representation of interior partner operators can be constructed by time-evolving a corresponding mode on the right hand side. This mode, constructed exclusively on the degrees of freedom on the right hand side, was called $R_{D}$. We now include the effect of the infalling observer $A$ as a gravitational shockwave and draw the backreacted geometry where the horizon is shifted as depicted in Fig.~\ref{fig-Penrose}(a). If the time separation between the outgoing mode $D$ and the infalling observer $A$ is longer than or equal to the scrambling time, the interior mode $\widetilde{D}$, which is distinct from $R_{D}$, can be found across the horizon. See~\cite{Beni19} for discussions on cases where the time separation is shorter than the scrambling time. 

One subtlety in this proposal is the necessity of $d_{A}\gtrapprox d_{D}$. For the purpose of resolving the firewall puzzle, namely as captured in the above AMPS thought experiment, this requirement can be easily justified. The outside observer handed distilled qubits $R_{D}$ to the infalling observer, so the infalling observer should carry more qubits than $R_{D}$ in her attempt to verify the violation of the monogamy of entanglement.

It is worth comparing our proposal with previous proposals, often bundled under the slogan ``$\text{ER}=\text{EPR}$''~\cite{Maldacena13}. These proposals roughly go as follows. The distillation of the qubit $R_{D}$ by the outside observer Bob creates perturbations which will become a high-energy density near the horizon as shown in Fig.~\ref{fig-Penrose}(b). This perturbation spoils the quantum entanglement between $D$ and $\widetilde{D}$ and creates a firewall (or Alice may be killed by the shockwave). As such, Alice is not able to observe entanglement between $D$ and $\widetilde{D}$. One may draw the Penrose diagram corresponding to these proposals by explicitly treating $R_D$ as a gravitational shockwave. Unfortunately, the decoupling theorem tells that scenarios along this line are not consistent with unitarity or scrambling dynamics of quantum black holes. 

Our proposed resolution of the firewall puzzle considers the effect of the perturbations from the infalling observer $A$ as a primary one, instead of $R_{D}$ as in the ``$\text{ER}=\text{EPR}$'' approaches, and draws the backreacted geometry with $A$ being treated as a gravitational shockwave. From the bulk perspective, it may not be immediately clear which perturbation, $A$ or $R_{D}$, should be taken into account as a primary source of backreaction. The important point here is that by relying on the prediction from the boundary quantum mechanical descriptions (namely, unitarity and OTOC decay) as a guiding principle, we can draw the Penrose diagram in an unambiguous manner (which is Fig.~\ref{fig-Penrose}(a), not Fig.~\ref{fig-Penrose}(b)). The decoupling theorem~\ref{theorem:decoupling} tells us that $R_{D}$ is not the primary source of perturbations on the bulk to describe the physics of the infalling observer.

Here the observer-dependence of the black hole interior plays a crucial role. The infalling observer will see an interior which is prepared by her own backreaction. Her infalling experience as well as her black hole interior will not be affected by any quantum operations localized on the early radiation. In fact, whether the black hole is young or old does not matter for crossing the horizon.

Scrambling phenomena has to do with entanglement dynamics which can be measured either by preparing two entangled copies of the same system or by evolving the system forward and backwards in time. It is quite interesting that the firewall problem can be resolved in a way which crucially relies on entanglement dynamics, but does not appear to involve time-reversal or two copies of the system. The key to understand this question is that the infalling observer, not the outside observer, will be the one to measure quantum entanglement by seeing the smoothness of the horizon. To communicate her infalling experience back to the outside observer, she needs to be pulled back to the outside. In our construction of interior operators, this was done by performing the EPR distillation. We may use the deterministic variant from~\cite{Yoshida:2017aa} if we want to save her for sure. The procedure does involve time-reversal which is required for verification of entanglement from the perspective of the outside observer. 

\subsection{Marolf-Wall puzzle}

\begin{figure}
\centering
\includegraphics[width=0.35\textwidth]{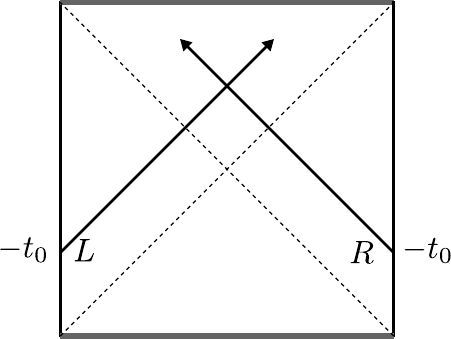}
\caption{Classical trajectories from the left and right sides can intersect inside the black hole. Can two observers meet?
}
\label{fig-meeting}
\end{figure}

Observer-dependence due to gravitational backreaction also provides a possible resolution of another interesting puzzle concerning the Hilbert space structure of two-sided black holes. Consider the two-sided AdS black hole geometry which is usually interpreted as the thermofield double state of the left and right CFTs whose Hilbert space has a factorized form $\mathcal{H}=\mathcal{H}_{L}\otimes \mathcal{H}_{R}$. In a naive semi-classical treatment on a fixed background geometry, it appears that two observers from the left and the right can be arranged to meet inside the black hole if they depart sufficiently before $t=0$. See Fig.~\ref{fig-meeting}. But the CFT Hamiltonian $H_{L}+H_{R}$ does not couple two sides, so two observers should not be able to meet. How do we make sense of these? 

A possible resolution of this puzzle was suggested by Marolf and Wall~\cite{Marolf:2012aa}. This puzzle itself was referred to as the Marolf-Wall puzzle in~\cite{Harlow:2016ab}. Their proposal was that we need additional degrees of freedom beyond the original CFTs, some sort of superselection sectors, in order to describe the meeting. This proposal is closely related to the tensor factorization problem as pointed out in~\cite{Harlow:2016aa}. 

Here, however, we arrive at a different resolution of the puzzle by considering the effect of backreaction. In fact, the structure of the problem is identical to the one in the firewall puzzle. Suppose that two signals were sent at $t= - t_{0}$ so that their semi-classical trajectories would intersect in the black hole interior (Fig.~\ref{fig-meeting}). For definiteness, we focus on the left signal $L$ as the one we start with and ask if it receives the right signal $R$ in the interior. In the AdS/CFT correspondence, the right signal $R$ can be interpreted as an excitation on the right CFT. Then, in the absence of the left signal, one can find some mode $R'$ on the left side at time $t = + t_{0}$ which is entangled with the mode $R$. In the unperturbed geometry, the right signal, started at $t=-t_{0}$, would travel along the mode $R$ which is realized as the entangled partner of $R'$. However, due to the backreaction by the left signal $L$, the near-horizon geometry will be shifted in the same manner as in the firewall problem. From the reasoning similar to section~\ref{sec:firewall}, we see that the right signal will not reach the left signal if $t_{0}\gtrapprox t_{\text{scr}}$. The original entangled partner mode $R$, where the right signal was supposed to travel along, is no longer entangled with the left side $R'$ due to the backreaction. Inside the black hole, the left signal $L$ will come across the newly created partner mode $\widetilde{R}$ where the right signal is absent. If $t_{0}\ll t_{\text{scr}}$, two signals would meet near the singularity where we should doubt the validity of semiclassical treatment from the beginning. 

The central idea behind this resolution is again that the black hole interior is observer-dependent. Left and right observers jump into their own black hole interiors which are separated degrees of freedom. 
Given a possible relation between the Marolf-Wall puzzle and the tensor factorization problem, we speculate that observer-dependence of black hole interiors may also provide useful insights on the latter problem. 

\subsection{Page curve and state-dependence}

Finally we discuss the implication of our state-independent reconstruction to the information loss problem. There are many variants of the black hole information puzzle, but we shall focus on the particular question of why the entanglement entropy starts to decrease after the Page time~\cite{Page:1993aa}. 

In the literature we often see that the Page curve problem and the firewall problem are discussed on the equal footing as if these have the same origin. While both problems are concerned about entangled partners of outgoing modes, their resolutions come from entanglement properties of completely opposite nature as we clarify below. 

The initial increase of the entanglement entropy results from the fact that the outgoing mode is entangled with the remaining black hole degrees of freedom. After the Page time, however the situation will change drastically. If unitarity is assumed from the outside observer's description, the outgoing mode after the Page time must be somehow entangled with the early radiation since otherwise the entanglement entropy will not decrease. This observation seems to suggest that the interior partner of the late outgoing mode must be constructed by using the early radiation degrees of freedom and hence must be state-dependent. It appears to be in tension with our view that interior partner operators can be reconstructed in a state-independent manner. 

Our proposal to resolve this apparent tension is that in an evaporating black hole, a large portion of outgoing modes can be still reconstructed in a state-independent manner, but a tiny portion accounting for the black hole evaporation must be state-dependent. 

To warm up, let us recap the interior reconstruction problem for a large AdS black hole which does not evaporate. The time-evolution of this quantum system can be depicted as follows:
\begin{align}
\figbox{1.0}{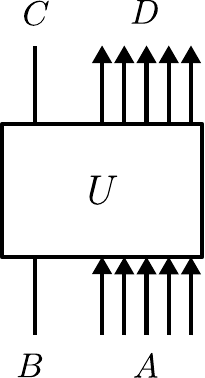}
\end{align}
where $A,D$ are infalling and outgoing modes respectively while $B,C$ are all the other degrees of freedom. State-independent reconstruction is possible as long as $d_{A} \gtrapprox d_{D}$. This condition is satisfied in the AdS black holes where the outgoing modes are reflected back at the boundary and hence $d_{A}=d_{D}$~\footnote{Strictly speaking, the reconstruction error scales as $O\Big(\frac{d_{D}^2}{d_{A}^2}\Big)$, so $A$ needs to contain a few more qubits than $D$. }.

For evaporating black holes, however, there will be more outgoing modes than infalling modes, corresponding to the situation with $d_{D} \gtrapprox d_{A}$. Let us examine how the state-independent reconstruction ceases to work when the subsystem $D$ becomes larger than $A$. A toy model which mimics this situation is depicted below:
\begin{align}
\figbox{1.0}{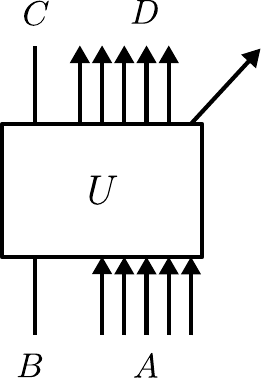}
\end{align}
where $|D| > |A|$. Small portion of the outgoing modes will escape to the environment while the rest returns to the black hole. For black holes in a flat space, those returning modes are typically confined in the radiation zone. 

Let us examine where $D$ is entangled with. 
When $U$ is scrambling and $d_{A}\leq d_{D}$, we obtain~\cite{Beni18}
\begin{align}
I(D,C\overline{A}) \approx 2 \log_{2} d_{A} \qquad I(D,\overline{B}) \approx 2 (\log_{2}d_{D} - \log_{2} d_{A}).
\end{align}
This suggests that there exists $\simeq d_{A}$-dimensional subspace ($\simeq \log_2 d_{A}$ qubits) of $D$ which is entangled with $C\overline{A}$. For operators acting on this subspace, entangled partners can be reconstructed on $C\overline{A}$ in a state-independent manner as these are not entangled with $\overline{B}$. However the remaining $\approx d_{D}/d_{A}$-dimensional subspace ($\simeq \log_2 d_{D} - \log_2 d_{A}$ qubits) are entangled with the early radiation $\overline{B}$. Hence, their entangled partners must be reconstructed in a state-dependent manner.

This subtle distinction between state-dependent and state-independent partner operators provides an interesting insight on the emergence of the Page curve in the evaporating black hole. Imagine that the outgoing mode is made of $\log_2 d_{D}$ qubits and release them one by one to the environment which joins the early radiation $\overline{B}$. When we release the first $\log_2 d_{A}$ qubits, entanglement entropy between the black hole and the early radiation will increase since those qubits are entangled with $C\overline{A}$ (plus qubits in $D$ which have not been released). On the other hand, the remaining $\log_{2} d_{D} - \log_2 d_{A}$ qubits are entangled with the early radiation $\overline{B}$, so releasing them will decrease the entanglement entropy between the black hole and the environment, leading to the Page curve behavior~\footnote{A similar argument for the state-dependence in the Page curve has been recently made in~\cite{Penington19}. In our understanding, however, we disagree on its implication to the firewall problem.}. 

The smoothness of the horizon requires the interior partners to be state-independent and the emergence of the Page curve requires those to be state-dependent. In an evaporating black hole, two types of interior operators can coexist. In the firewall problem, we typically focus on situations where state-independent modes are dominant as evaporation process is assumed to be slow.

\section*{Acknowledgment}

I thank Yoni BenTov and Yingfei Gu for discussions.
Research at the Perimeter Institute is supported by the Government of Canada through Innovation, Science and Economic Development Canada and by the Province of Ontario through the Ministry of Economic Development, Job Creation and Trade.

\appendix

\section{Review of quantum information concepts}\label{sec:review}

\subsection{Scrambling}

Here we recall the definition of scrambling in terms of OTOCs. 

At infinite temperature, OTOCs are defined with respect to the maximally mixed state $\rho=\frac{1}{d}\mathbb{I}$:
\begin{align}
\langle W(t) Z(0) Y(t) X(0) \rangle \equiv \frac{1}{d}\Tr [ W(t) Z(0) Y(t) X(0)]
\end{align}
where time-evolved operators are defined by $O(t)=U^{\dagger}O(0)U$. The phenomena of scrambling is often associated with the decay (decorrelation) of OTOCs from their initial values when $W,Z,Y,X$ are basis operators such as Pauli and Majorana operators. In a more genetic language, the system is said to be scrambling at time $t$ if the following asymptotic decomposition holds for all the local operators $W,Z,Y,X$~\cite{Yoshida:2017aa}:
\begin{align}
\langle W(t) Z(0) Y(t) X(0) \rangle 
\approx \langle WY \rangle \langle Z \rangle \langle X \rangle
+ \langle ZX \rangle \langle W \rangle \langle Y \rangle - \langle W \rangle \langle Z \rangle\langle Y \rangle\langle X \rangle \label{eq:scrambling-definition}
\end{align}
where expectation values are defined by $\langle O \rangle \equiv \frac{1}{d}\Tr[O]$. 

We immediately see that the asymptotic value for a large system size is zero when $X,Y,Z,W$ are traceless. This asymptotic form has been derived for Haar random unitary~\cite{Roberts:2017aa} and from the Eigenstate Thermalization Hypothesis (ETH)~\cite{Huang:2019aa}. Corrections to the asymptotic form are polynomially (exponentially) suppressed with respect to the system size $n$ for systems with (without) conserved quantities. 

At finite temperature, OTOCs are defined with respect to a thermal mixed state $\rho = e^{-\beta H}/\Tr (e^{-\beta H}) $:
\begin{align}
\langle W(t) Z(0) Y(t) X(0) \rangle \equiv \Tr [\rho^{\alpha} W(t) \rho^{\beta}Z(0)\rho^{\gamma} Y(t)\rho^{\delta} X(0)] \label{eq:exponent}
\end{align}
where 
\begin{align}
\alpha,\beta,\gamma,\delta \geq 0 \qquad \alpha + \beta + \gamma + \delta = 1.
\end{align}
Specifics of $\alpha,\beta,\gamma,\delta$ do not affect the main result, and we will focus on $\alpha=\beta=\gamma=\delta = 1/4$ unless otherwise stated. We can infer the asymptotic form of finite temperature OTOCs in an analogous manner by using thermal expectation values:
\begin{align}
\langle O \rangle \equiv \Tr(\rho O) \qquad \langle O_{1}O_{2} \rangle \equiv \Tr(\rho^{\alpha} O_{1} \rho^{\beta} O_{2}) \qquad \alpha,\beta>0\qquad \alpha+\beta =1.
\end{align}
Again, specifics of $\alpha,\beta$ does not affect the main result, and we will choose $\alpha=\beta=1/2$. The asymptotic form of OTOCs at finite temperature can be also derived from ETH~\cite{Alvaro_Private}.

Let us present some intuition behind the aforementioned definition of scrambling. The thermal ensemble $\rho$ has significant contributions from eigenstates within typical energy subspace at given temperature. Hence one may interpret each insertion of powers of $\rho$ as projection onto typical energy subspace. The asymptotic form of OTOC suggests that the time-evolution $U$ resembles Haar random unitary within the typical energy subspace. Here we are particularly interested in operators $X,Y,Z,W$ which do not change the energy of the system much so that the system remains within the typical energy subspace. 

It is worth emphasizing that our main focus is on systems after the scrambling time when OTOCs \emph{have} decayed to asymptotic values. The famous Lyapunov behavior is typically observed only before the scrambling time.

\subsection{Diagrams}

In this paper, we extensively use diagrammatic tensor notations in order to express wavefunctions and operators as well as physical processes. While these techniques have become standard in the past several years, it would be still beneficial to have a brief summary to make the presentation self-consistent. Wavefunctions and operators are represented by 
\begin{align}
|\psi\rangle \ = \ \figbox{1.0}{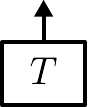} \qquad \langle \psi| = \figbox{1.0}{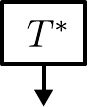}
\qquad O \ =\ \figbox{1.0}{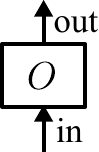}
\end{align}
which can be also explicitly written as
\begin{align}
|\psi\rangle = \sum_{j} T_{j}|j\rangle \qquad \langle \psi |\ = \ \sum_{j} T_{j}^* \langle j | \qquad O\ = \  \sum_{ij} O_{ij}|i\rangle \langle j |.
\end{align}
By using these tensors as building blocks, one can express various physical processes in a graphical manner. For instance, an expectation value can be represented by
\begin{align}
\langle \psi | O_{n}\cdots O_{1} |\psi\rangle \ =\ \figbox{1.0}{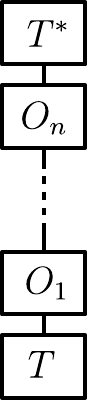}\ .
\end{align}
In order to associate a physical process to an equation like $\langle \psi | O_{n}\cdots O_{1} |\psi\rangle$, one needs to read it from the right to the left, \emph{i.e.} the initial state $|\psi\rangle$ is acted by $O_{1},O_{2},\cdots$ sequentially and then is projected onto $|\psi\rangle$. In the diagrammatic notation, one needs to read the figure from the bottom to the top, \emph{i.e.} the time flows upward in the diagram.

Consider a Hilbert space $\mathcal{H}_{R} = \mathbb{C}^{d_{R}}$ whose dimensionality is $d_{R}\equiv|\mathcal{H}_{R}|$. An identity operator, $I_{R} = \sum_{j=1}^{d_{R}} |j\rangle \langle j |$, can be expressed as a straight line (\emph{i.e.} a trivial tensor) since its inputs and outputs are the same:
\begin{align}
I_{R} \ = \ \figbox{1.0}{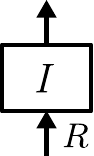}  \ = \ \figbox{1.0}{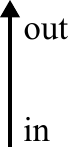}\ .
\end{align}
This diagram has one input leg (index) and one output leg. One may bend the line and construct the following diagram:
\begin{align}
|\text{EPR}\rangle_{R\overline{R}} \propto \sum_{j=1}^{d_{R}} |j\rangle \otimes |j\rangle  \ = \ \figbox{1.0}{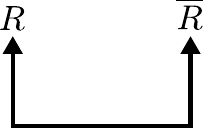}
\end{align}
which is the same trivial line, but with two output legs instead of one in and one out. This diagram represents an unnormalized EPR pair defined on $\mathcal{H}_{R}\otimes \mathcal{H}_{\overline{R}}$. 

We utilize the following notation for normalized EPR pairs
\begin{align}
|\text{EPR}\rangle_{R\overline{R}} = \frac{1}{\sqrt{d_{R}}} \sum_{j=1}^{d_{R}} |j\rangle \otimes |j\rangle  \ = \ \figbox{1.0}{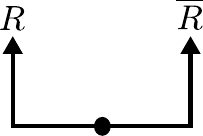}
\end{align}
where the black dot carries a factor of $1/\sqrt{d_{R}}$. 

\subsection{$O, O^{\dagger}, O^T, O^*$}

It is also beneficial to define $O, O^{\dagger}, O^T, O^*$ and review some useful relations associated with them. Given an operator $O$, the complex-conjugate transpose $O^{\dagger}$, the transpose $O^T$ and the complex-conjugate are defined by
\begin{align}
O = \sum_{i,j}O_{ij}|i\rangle \langle j| \qquad 
O^{\dagger} = \sum_{i,j}O_{ij}^*|j\rangle \langle i| \qquad 
O^{T} = \sum_{i,j}O_{ij}|j\rangle \langle i| \qquad 
O^* = \sum_{i,j}O_{ij}^*|i\rangle \langle j|
\end{align}
where $O^\dagger = (O^T)^*$. In the tensor diagrams, the transpose can be represented as follows:
\begin{align}
\figbox{1.0}{fig-notation-O}\ = \ \figbox{1.0}{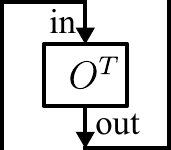}\ .
\end{align}
The following relation is particularly useful in constructing partner operators:
\begin{align}
(O \otimes I )|\text{EPR}\rangle = (I \otimes O^T )|\text{EPR}\rangle 
\end{align}
or graphically
\begin{align}
\figbox{1.0}{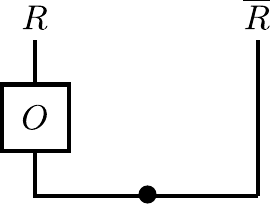}\ = \ \figbox{1.0}{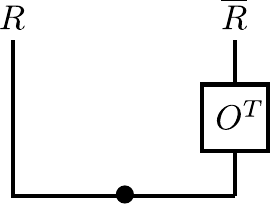}\ .
\end{align}

\subsection{State representation}

Given an operator $O$ acting on a Hilbert space $\mathcal{H}$, its state representation (Choi representation), supported on $\mathcal{H}\otimes \mathcal{H}$, is defined by
\begin{align}
|O\rangle \equiv (O\otimes I) |\text{EPR}\rangle =  \ \figbox{1.0}{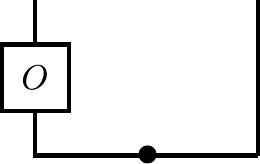} \ .
\end{align}
In this regard, the EPR pair is a state representation of an identity operator; $|\text{EPR}\rangle = |I\rangle$. See~\cite{Hosur:2015ylk} for further discussions on the state representation. 

Given a density matrix $\rho_{R}$ on $\mathcal{H}_R$, consider a purification (thermofield double state) on $\mathcal{H}_{R}\otimes \mathcal{H}_{\overline{R}}$:
\begin{align}
\rho = \sum_{j}p_{j} |\psi_{j}\rangle\langle \psi_{j}|\ \longrightarrow \  \sum_{j}\sqrt{p_{j}} |\psi_{j}\rangle \otimes | \psi_{j}^* \rangle.
\end{align}
The purified state can be viewed as a state representation of $\rho_{R}$. By introducing an unnormalized state 
\begin{align}
\widetilde{\rho}_{R} \equiv d_{R} \rho_{R} \qquad \Tr[\widetilde{\rho}_{R}] = d_{R},
\end{align} 
we have
\begin{align}
| \widetilde{\rho}_{R}^{1/2}\rangle = \sum_{j}\sqrt{p_{j}} |\psi_{j}\rangle \otimes | \psi_{j}^* \rangle.
\end{align}

\section{Statement of decoupling theorem}\label{sec:decoupling-proof}

Below we sketch a proof of $\rho_{\overline{B}D}\approx \rho_{\overline{B}}\otimes \rho_{D}$. When two states are close in $L^1$ distance, they cannot be distinguished by any measurement. The strategy is to show that $\rho_{\overline{B}D}$ is close to a maximally mixed state $\frac{1}{d_{\overline{B}} d_{D}}I_{B}\otimes I_{D}$ in the $L^1$ distance. More precisely, we will prove the following bound:
\begin{align}
\Big|\Big|\rho_{\overline{B}D} - \frac{1}{d_{\overline{B}} d_{D}}I_{B}\otimes I_{D}\Big|\Big|_{1}^2 \leq O\left( \frac{d_{D}^2}{d_{A}^2} \right) + O(\epsilon)
\end{align}
where the $L^1$ norm is defined by $||O||_{1}=\Tr \sqrt{O^\dagger O}$. Here, $\epsilon$ is the error from the asymptotic form of OTOCs which is typically suppressed polynomially in the system size $n$. 

The derivation is straightforward. Observe
\begin{align}
\Big|\Big|\rho_{\overline{B}D} - \frac{1}{d_{\overline{B}} d_{D}}I_{B}\otimes I_{D}\Big|\Big|_{1}^2 \leq d_{\overline{B}} d_{D} \Big|\Big|\rho_{\overline{B}D} - \frac{1}{d_{\overline{B}} d_{D}}I_{B}\otimes I_{D}\Big|\Big|_{2}^2 = \Tr(\rho_{\overline{B}D}^2) d_{\overline{B}}d_{D} - 1.
\end{align}
The RHS can be evaluated by noticing
\begin{align}
\Tr(\rho_{\overline{B}D}^2) = \frac{d_{D}}{d_{\overline{B}}}\Delta . 
\end{align}
Using the OTOC asymptotics, we obtain the claimed bound.

A similar bound can be derived for the cases with a pure state input by analogous analysis. We obtain
\begin{align}
\Big|\Big|\rho_{\overline{B}D} - \frac{1}{d_{\overline{B}} d_{D}}I_{B}\otimes I_{D}\Big|\Big|_{1}^2 \leq O\left( \frac{d_{D}^2}{d_{A}} \right) + O(\epsilon).
\end{align} 

For a finite temperature generalization, see~\cite{Yoshida:2017aa} and appendix in~\cite{Yoshida:2019aa}.

\section{Finite temperature}\label{sec:finite}

We construct partner operators at finite temperature. We consider the following quantum state at finite temperature:
\begin{align}
|\Psi\rangle \equiv (U\otimes I_{\overline{A}\overline{B}}) (|\widetilde{\rho_{B}}^{1/2}\rangle_{B\overline{B}} |\widetilde{\rho_{A}}^{1/2}\rangle_{A\overline{A}}) = \ \figbox{1.0}{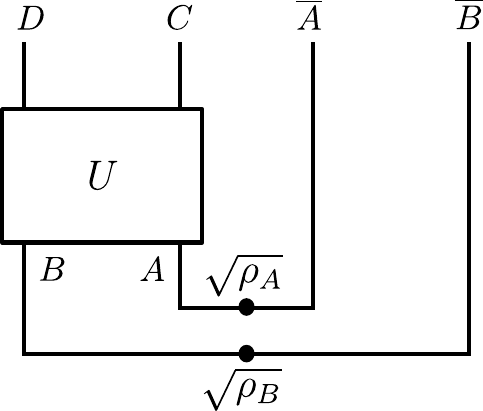}
\end{align}
where the black hole was initially in the thermofield double state $|\widetilde{\rho_{B}}^{1/2}\rangle$ on $B\overline{B}$, and then was perturbed by the measurement probe prepared in some entangled state $|\widetilde{\rho_{A}}^{1/2}\rangle$. 

The construction procedure be applied to finite temperature systems which satisfy a certain technical assumption~\cite{Yoshida:2017aa}. Given the input ensemble $\rho_{AB}=\rho_{A}\otimes \rho_{B}$, we will assume that the output ensemble factors into $C$ and $D$:
\begin{align}
\rho_{CD} \equiv U \rho_{AB}U^{\dagger} \approx \rho_{C}\otimes \rho_{D}.
\end{align}
Unfortunately, this condition is problematic for physical reasons; if $\rho_{CD}$ is thermal, $C$ and $D$ are entangled. We will present a justification for such an assumption in appendix~\ref{sec:justification}. For our purpose, it suffices to coarse-grain preexisting correlations between $C$ and $D$ in a thermal ensemble $\rho_{CD}$. 

\subsection{Distillation}

Having imposed the aforementioned assumption, generalization to finite-temperature systems is straightforward; we simply replace all the dots in the previous diagrammatic calculations with $\sqrt{\rho_{R}}$. One subtlety is that, since the outgoing mode $D$ is thermal, partner operators $\Omega(O_{D})$ must satisfy 
\begin{align}
\langle \Psi | O_{D}^{\dagger}\otimes \Omega(O_{D}) |\Psi \rangle \approx \langle O_{D}^\dagger O_{D} \rangle \equiv 
\Tr\Big[\rho_{D}^{1/2}  O_{D}^\dagger \rho_{D}^{1/2}  O_{D} \Big],
\end{align}
reproducing thermal expectation values. This relation can be viewed as the definition of interior partner operators for finite temperature systems. 

Hence we consider distillation protocol of $|\widetilde{\rho_{D}}^{1/2}\rangle$ from $D$ and $C\overline{A}$, which proceeds in the following manner:
\begin{align}
\Pi_{A\overline{A}} |\Phi_{\text{back}}\rangle = \ \figbox{1.0}{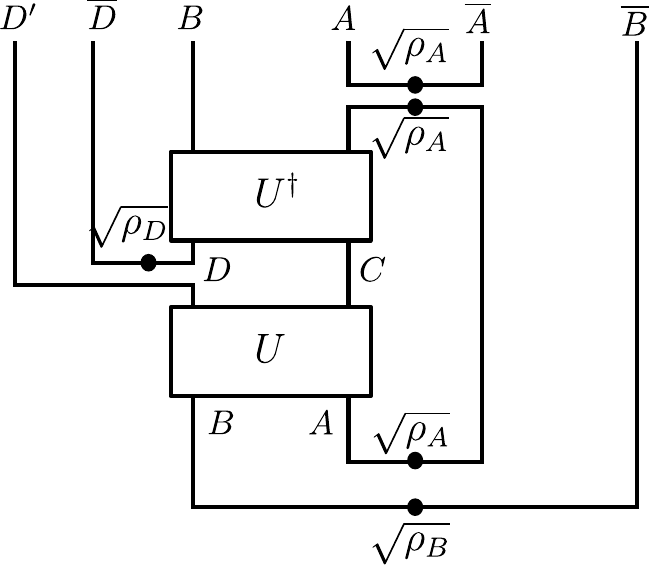}
\end{align}
where $|\widetilde{\rho_{D}}^{1/2}\rangle$ is inserted on $\overline{D}D$ and projection $\Pi_{A\overline{A}}\equiv|\widetilde{\rho_{A}}^{1/2}\rangle\langle \widetilde{\rho_{A}}^{1/2} |$ is applied on $A\overline{A}$.

The probability amplitude to measure $|\widetilde{\rho_{A}}^{1/2}\rangle$ can be evaluated via the asymptotic form of OTOCs at finite temperature. After some calculations (see appendix~\ref{sec:calculation}), we obtain
\begin{align}
\Delta \equiv \langle\Phi_{\text{back}} |\Pi_{A\overline{A}} |\Phi_{\text{back}}\rangle \approx \Tr[\rho_{D}^3]
\end{align}
for $d_A\gg d_D$~\footnote{This is the result for $\alpha=\beta=\gamma=\delta=1/4$ in Eq.~\eqref{eq:exponent}. When $\alpha=1$, we obtain $\Delta \approx \Tr[\rho_{D}^3] \Tr[\rho_{A}^{3/2}]\Tr[\rho_{A}^{1/2}]$. If the variance of the spectrum of $\rho_{A}$, normalized by its mean, approaches zero asymptotically, $\Tr[\rho_{A}^{3/2}]\Tr[\rho_{A}^{1/2}]\approx 1$. }. 
The following lower bound is useful:
\begin{align}
\langle\Phi_{\text{back}}| \Pi_{D'\overline{D}} \Pi_{A\overline{A}} |\Phi_{\text{back}}\rangle \geq \langle\Phi_{\text{back}}| \Pi_{D'\overline{D}} \Pi_{A\overline{A}} \Pi_{B\overline{B}}|\Phi_{\text{back}}\rangle = \Tr[\rho_{D}^2]^2.
\end{align}
Hence, the probability amplitude to measure $|\widetilde{\rho_{D}}^{1/2}\rangle$ after the postselection is lower bounded by
\begin{align}
\frac{1}{\Delta}\langle\Phi_{\text{back}} | \Pi_{D'\overline{D}} \Pi_{A\overline{A}} |\Phi_{\text{back}}\rangle \gtrapprox \frac{\Tr[\rho_{D}^2]^2}{\Tr[\rho_{D}^3]} = 2^{2(S_{D}^{(3)} - S_{D}^{(2)} )}
\end{align}
where $S_{D}^{(\alpha)}$ is the R\'{e}nyi-$\alpha$ entropy of $\rho_{D}$. If $S_{D}^{(3)} - S_{D}^{(2)}\approx 0$, $|\widetilde{\rho_{D}}^{1/2}\rangle$ can be distilled on $D'\overline{D}$. 

\subsection{Dressed operator growth}\label{sec:growth-T}

The distillation procedure constructs the following interior partner operators:
\begin{align}
\Omega(O_{D})= \ \figbox{1.0}{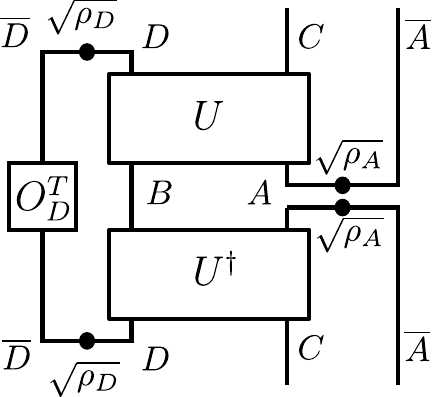}\ .
\end{align}
Let us decompose $\Pi_{A\overline{A}}=|\widetilde{\rho_{A}}^{1/2}  \rangle\langle \widetilde{\rho_{A}}^{1/2} |$ in terms of basis operators (such as Pauli operators):
\begin{align}
\Pi_{A\overline{A}} = \sum_{V,W} \alpha_{V,W} V_{A} \otimes W_{\overline{A}} \qquad \alpha_{V,W} = \langle \widetilde{\rho_{A}}^{1/2} | V_{A}^{\dagger} \otimes W_{\overline{A}}^{\dagger} | \widetilde{\rho_{A}}^{1/2}  \rangle. 
\end{align}
Each term in the decomposition contributes to $\Omega(O)$ as follows:
\begin{align}
 \figbox{1.0}{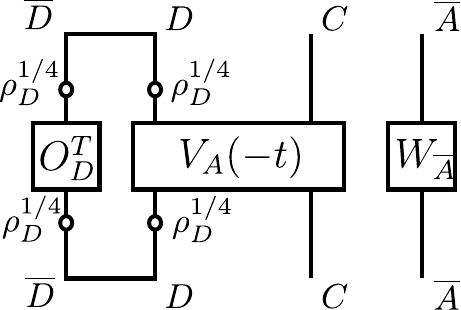}\ \label{eq:thermal-collision}
\end{align}
Here it is convenient to consider thermally dressed operators~\cite{Qi:2019aa}:
\begin{align}
\widetilde{O_{R}} \equiv  \widetilde{\rho_{R}}^{1/4}O_{R} \widetilde{\rho_{R}}^{1/4}
\end{align}
Then the diagram in Eq.~\eqref{eq:thermal-collision} can be interpreted as truncation of $\widetilde{V_{A}}^{\dagger}(-t)$ by $\widetilde{O_{D}}^{\dagger}$. Hence, we see that interior partner operators can be constructed due to the thermally dressed operator growth.

\section{OTOC calculation}\label{sec:calculation}

Let us represent OTOCs graphically as follows:
\begin{align}
\text{OTOC}(P,Q) \equiv \figbox{1.0}{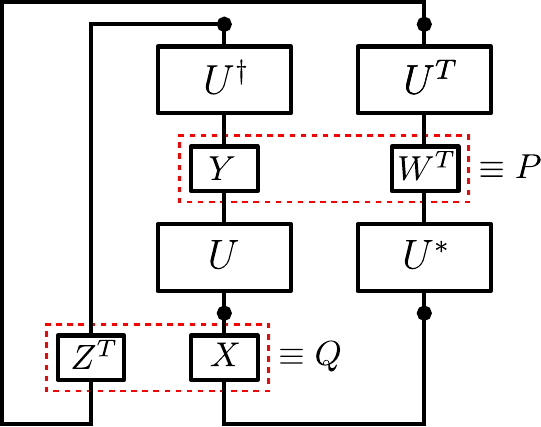}  \qquad P = Y \otimes W^{T} \qquad Q = Z^{T}\otimes X.
\end{align}
Due to linearity, we can generalize the above OTOCs as a function of $P,Q$ which act on $\mathcal{H}^{\otimes 2}$. We then have
\begin{align}
\Delta = \text{OTOC}(P,Q) \qquad P = \figbox{1.0}{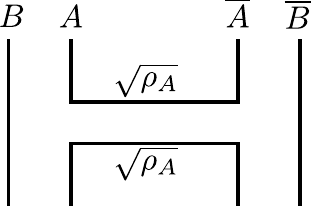} \qquad Q = \figbox{1.0}{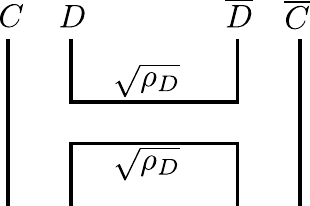}\ .
\end{align}
Using the OTOC asymptotics, we obtain
\begin{align}
\Delta = \Tr(\rho_{D}^3) + \Tr(\rho_{A}^3) -   \Tr(\rho_{D}^3) \Tr(\rho_{A}^3) \approx \Tr(\rho_{D}^3).
\end{align}

\section{On $\rho_{CD}\approx \rho_{C}\otimes \rho_{D}$}\label{sec:justification}

Here we present a justification of the assumption $\rho_{CD}\approx \rho_{C}\otimes \rho_{D}$~\footnote{I thank Alexei Kitaev for discussions on this.}. 

\subsection{Coarse-graining}

For simplicity of discussion, we approximate $\rho_{CD}$ as a thermal ensemble $\rho_{CD}=\frac{e^{-\beta H_{CD}}}{\Tr[ e^{-\beta H_{CD}} ]}$. Consider the thermofield double state $|\widetilde{\rho_{CD}}^{1/2}\rangle$ supported on $C,D,\overline{C},\overline{D}$. The assumption $\rho_{CD}\approx \rho_{C}\otimes \rho_{D}$ asserts that $C$ ($D$) is entangled only with $\overline{C}$ ($\overline{D}$) respectively. In general, however, this is not the case. 

\begin{figure}
\centering
\includegraphics[width=0.45\textwidth]{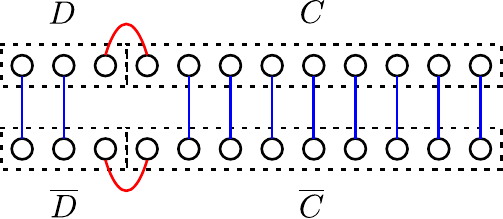}
\caption{Short-range entanglement between $C$ and $D$. 
}
\label{fig-SRE}
\end{figure}

Here we are interested in what degrees of freedom the outgoing mode $D$ are entangled with. 
Eigenstates of the Hamiltonian $H_{CD}$ are entangled between $C$ and $D$. At low temperature, a fewer number of states contribute to the thermal ensemble and thus entanglement between $C$ and $D$ survives in $\rho_{CD}$. Then the ``partner'' operators of $D$ can be found on $C$ due to the preexisting entanglement in $\rho_{CD}$. In this case, the system does not need to be perturbed in order to construct the partner operator from the beginning. Indeed, such modes on $D$ do not lead to the firewall puzzle since, in order for the puzzle to exist, $D$ needs to be entangled with $\overline{C}\overline{D}$ in the first place. See Fig.~\ref{fig-SRE}.

There is another reason why entanglement between $C$ and $D$ may not be an interesting one to focus on. In discrete spin systems, entanglement between $C$ and $D$ are mostly of short-range nature. In quantum field theories, such entanglement is UV divergent and hence is cut-off dependent property which should be removed by some proper coarse-graining. We may simply discard short-range entanglement between $C$ and $D$ and construct subspace $C_{sub}$ and $D_{sub}$ on which we may have a factor structure $\rho_{C_{sub}}\otimes \rho_{D_{sub}}$. Due to the preexisting entanglement between $C$ and $D$, releasing $D$ to the environment appears to increase entanglement entropy of the black hole. These UV divergent pieces, however, are often ignored in the Bekenstein bound~\cite{Casini:2008aa}. 

\subsection{Soft mode}

Even if we remove short-range entanglement between $C$ and $D$, however, there still remains some classical correlation between $C$ and $D$ due to the presence of conserved quantities (such as the energy, charge and angular momentum). The Hayden-Preskill recovery problem for systems with conserved quantities (symmetries) has been discussed in~\cite{Beni18}. Here it is convenient to consider two types of physical modes (hard and soft modes) on $C,D$. Roughly, hard modes are those which are associated with change of conserved quantities on $C,D$ whereas soft modes are those which do not change them. It is convenient to imagine an approximate decomposition of the Hilbert space into a block-diagonal form: $\mathcal{H}\approx \bigoplus_{\alpha} \mathcal{H}^{\alpha}_{C} \otimes \mathcal{H}^{\alpha}_{D}$ where $\alpha$ labels the conserved quantity. Soft mode operators are block diagonal while hard mode operators may change $\alpha$-labels by acting on both $C$ and $D$. Here the thermal density matrix $\rho_{\beta}$ can be interpreted as an approximate projection onto states within a typical energy window. Such a subspace corresponds to some single label of $\alpha$ in the above decomposition, and soft mode operators do not change the energy of the system much. Examples of soft modes include gravitational shockwaves which only shift the horizon, commuting with $\rho_{\beta}$. When $\alpha$-label in $C$ changes, $\alpha$-label in $D$ must change too due to the conservation law. This suggests that hard modes on $D$ are correlated classically with hard modes on $C$. Hence we can focus on entanglement associated with soft modes only. In~\cite{Beni18}, it has been shown that soft modes on $D$ can be reconstructed by using the Hayden-Preskill recovery protocols.

\providecommand{\href}[2]{#2}\begingroup\raggedright\endgroup

\end{document}